\documentclass[a4paper,12pt]{article}
\usepackage[utf8]{inputenc}
 \usepackage[margin=1.1in]{geometry}
 \usepackage{amssymb} 
\usepackage{amsmath}
\usepackage{authblk}
\usepackage{comment}
\usepackage{tikz}
\usepackage{graphicx}
\usepackage{array}
\usepackage[hyperindex,breaklinks]{hyperref}
\usepackage[numbers,sort&compress]{natbib}
\usepackage{braket}

\title{\huge Vacuum structure and electroweak phase transition in singlet scalar dark matter}
\author{\vspace*{2cm}Parsa Ghorbani}
\affil{
\normalsize \it Dipartimento di Fisica dell'Universit{\`a} di Pisa, Italy \\
\normalsize \it INFN, Sezione di Pisa, Italy\\
}
\date{}

\begin{document}
\maketitle

\begin{abstract}
In the presence of a real singlet scalar field with $\mathbb{Z}_2$ symmetry in addition to the Higgs field in the Standard Model, we analytically investigate all possible one-step and two-step electroweak phase transitions (EWPT) in the high-temperature expansion limit. In particular, we examine the possibility of a first-order phase transition in an intermediate temperature interval of the universe for two-step scenarios. In all scenarios, we provide with necessary conditions on the parameters of the model to address a first- or second-order phase transition if ever possible in the high-temperature approximation. We show that among all possible EWPT channels in this model, only the two-step phase transition $(0,0)\to(0,w)\to (v,0)$ can be of first-order type, either in the first or second step. The conditions obtained for EWPT is confronted with the dark matter (DM) and collider constraints. It is shown that the second-order one-step $(0,0)\to(v,0)$ scenario can evade the DM constraints, but it is not accessible to future colliders. The two-step first-order EWPT scenario $(0,0)\to(0,w)\to (v,0)$, except for small DM masses, is excluded by direct detection experiments XENON1T/LUX, but it is accessible in future colliders ILC/TLEP.
\end{abstract}

\newpage

\section{Introduction}

The vacuum structure in the Standard Model (SM) with the quartic Higgs potential at finite temperature is fairly well understood \cite{Bochkarev:1987wf,Dine:1992wr}. It is known that the SM with the Higgs mass fixed at $m_H\simeq 126$ GeV cannot undergo a strong first-order phase transition from the symmetric phase with vanishing Higgs vacuum expectation value (VEV), to the broken phase where the electroweak gauge group $SU(2)_L\times U(1)_Y$ breaks down due to the Higgs non-zero VEV. When we add more scalars to the SM, the vacuum structure becomes more complicated rapidly with the number of the scalar degrees of freedom (see \cite{Ghorbani:2019itr,Carena:2019une,Kannike:2020qtw} for the vacuum structure in the level of two additional real singlet scalars). 

Electroweak phase transition beyond the SM with only one extra real singlet scalar has been elaborately studied in the literature \cite{Espinosa:1993bs,Choi:1993cv,Profumo:2007wc,Ahriche:2007jp,Ashoorioon:2009nf,Espinosa:2011ax,Cline:2012hg,Kozaczuk:2015owa,Damgaard:2015con,Ghosh:2015apa,Beniwal:2017eik,Chiang:2018gsn}
Already adding only one single neutral scalar to the SM provides different vacuum solutions for which various phase transition can be envisioned. Some of these scenarios have been studied in the literature \cite{Espinosa:2011ax,Vaskonen:2016yiu,Ghorbani:2018yfr}. However, it is not investigated whether other channels of first-order phase transition is possible. Specially it is not known whether or not a first-order electroweak phase transition within an intermediate temperature interval in the early universe followed by a first- or second-order transition in a second step down to zero temperature is possible.  
In this paper we try to answer this question with the simplest model beyond the SM, namely considering an extension of the SM with a real singlet scalar denoted by $s$ with the $\mathbb{Z}_2$ symmetry,
\begin{equation}\label{3levelpot}
 \mathcal{V}_0(H,s)=-\mu^2_\text{h} |H|^2+\lambda_\text{h} |H|^4-\frac{1}{2}\mu_\text{s}^2 s^2 +\frac{1}{4}\lambda_\text{s} s^4 + \lambda_\text{hs} |H|^2 s^2 \,.
\end{equation}

In the high-temperature expansion limit where $T\gg {m_i}$ with $m_i$ being the mass of particles in the theory, the leading thermal corrections are proportional to $T^2$ while the subleading linear and logarithmic temperature dependent terms as well as the Coleman-Weinberg one-loop zero temperature corrections are negligible and can be omitted from the thermal effective potential \cite{Arnold:1992rz}. Therefore, the thermal effective potential can be written as (see e.g. \cite{Quiros:2007zz,Senaha:2020mop}), 
\begin{equation}
 \mathcal{V}_\text{eff}^\text{1-loop}(h,s;T) \approx \mathcal{V}_0(h,s)+\sum_{\phi_B,\phi_F} \left( \frac{g_B}{12}  m^2(\phi_B)+\frac{g_F}{48} m^2(\phi_F) \right) T^2
\end{equation}
where $\phi_B$ ($\phi_F$) consists of all bosonic (fermionic) fields with  $g_B$ ($g_F$) being the bosonic (fermionic) degrees of freedom and $m(\phi_B)$ ($m(\phi_F)$) denotes the mass of the bosonic (fermionic) field. After gauging away three components of the Higgs doublet we are left with the neutral Higgs component $h$; $H^\dagger=(0~h/\sqrt{2})$. For the real singlet scalar extension of the SM, the bosonic fields consist of the gauge bosons, $W, Z$, the Higgs field $h$ and the extra real scalar $s$. The most contribution of the fermionic fields comes from the top quark $t$ (neglecting light quarks). Therefore, the field-dependent masses are given by,
\begin{equation}
 \begin{split}
 & m^2_t(h)=y^2_t h^2\\
 &m^2_W(h)=g^2 h^2/4\\
  & m^2_Z(h)=(g^2+g'^2)h^2/4\\
  & m^2_\text{h}(h,s)=-\mu^2_\text{h}+3\lambda_\text{h} h^2+\lambda_\text{hs} s^2\\
 & m^2_\text{s}(h,s)=-\mu^2_\text{s}+3\lambda_\text{s} s^2+\lambda_\text{hs} h^2\,.
 \end{split}
\end{equation}
Taking into account the number of degrees of freedom for each field, the temperature-dependent part of the thermal effective potential in the high-temperature limit (neglecting the zero temperature one-loop effects) reads, 
\begin{equation}
 \mathcal{V}_\text{eff}^\text{1-loop}(h,s;T)\supset \frac{1}{2}\left( c_\text{h} h^2+c_\text{s} s^2 \right) T^2
\end{equation}
with 
\begin{subequations}
\begin{align}
 &c_\text{h}= \frac{1}{48}(9g^2+3g'^2+12y^2_t+12 \lambda_\text{h}+4\lambda_\text{hs}) \\
 &c_\text{s}= \frac{1}{12}(3\lambda_\text{s}+\lambda_\text{hs})
\end{align}
where the values of the couplings in the SM are known: $g\simeq0.65, g'\simeq0.35, \lambda_\text{h}\simeq0.129$ and $y_t\simeq1$. In other words, $c_\text{h}\simeq 0.37+\lambda_\text{hs}/12$ and $c_\text{s}=\lambda_\text{s}/4+\lambda_\text{hs}/12$ for the singlet scalar model.
\end{subequations}
This is equivalent to replace the parameters in the potential Eq. (\ref{3levelpot}) by the temperature-dependent parameters, 
\begin{equation}\label{muT}
 \mu^2_\text{h}\to\mu^2_\text{h}(T)\equiv \mu^2_\text{h}-c_\text{h} T^2  ~~~~~ \mu^2_\text{s}\to\mu^2_\text{s}(T)\equiv \mu^2_\text{s}-c_\text{s} T^2
\end{equation}
so that the thermal effective potential then takes the following form, 
\begin{equation}\label{therpot}
  \mathcal{V}_\text{eff}(h,s;T)=-\frac{1}{2}\mu^2_\text{h}(T) h^2+\frac{1}{4}\lambda_\text{h} h^4-\frac{1}{2}\mu_\text{s}^2(T) s^2 +\frac{1}{4}\lambda_\text{s} s^4 + \frac{1}{2}\lambda_\text{hs} h^2 s^2 \,.
\end{equation}
When the extra scalar sector possesses the $\mathbb{Z}_2$ symmetry, at very high temperature, the only possible extremum of the model is the $(0,0)$ solution. The minimum condition for that requires, 
\begin{equation}\label{chcs}
 c_\text{h}>0 ~\wedge~ c_\text{s}>0
\end{equation}
for all scenarios we study in this paper. 

This model apart from the $(0,0)$ solution admits three more vacuum solutions  $(v,0)$, $(0,w)$ and $(v,w)$ which break either the electroweak symmetry group or the $\mathbb{Z}_2$ discrete symmetry or both. The following constraints are taken into account in the next sections:

$\bullet$ The VEVs $v$ and $w$ in the extremum solutions throughout the paper are real and in general functions of the temperature $T$ and the parameters of the potential in Eq. (\ref{3levelpot}). 

$\bullet$ The existence of a real and positive critical temperature $T_c$   puts constraints on the parameters. 

$\bullet$ The field- and temperature-dependent Hessian matrix can be written as 
\begin{equation}\label{hessian}
\mathcal{H}(h,s;T)=\left(
 \begin{matrix}
  3\lambda_\text{h}  h^2 +  \lambda_\text{hs} s^2- \mu^2_\text{h}+c_\text{h} T^2  & 2  \lambda_\text{hs}h s\\
 2 \lambda_\text{hs} h s  &   \lambda_\text{hs}h^2 + 3  \lambda_\text{s} s^2- \mu^2_\text{s}+c_\text{s} T^2\\
 \end{matrix}\right)
\end{equation}
which will be used in the next sections for determination of the minima for different extremum solutions. 
Since the Hessian matrix in Eq. (\ref{hessian}) is symmetric, a generic extremum solution $(v,w)$ at a given temperature is a local minimum if $\mathcal{H}_{11}>0$ and $\text{Det}\mathcal{H}>0$\footnote{Note that the minimum condition is sufficient but not necessary. When $\text{Det}\mathcal{H}=0$ the second-derivative test is inconclusive.}. Equivalently, a generic extremum solution $(v,w)$ is a minimum if all mass eigenvalues are non-tachyonic. We will use either of these criteria in the next sections.   

$\bullet$ The positivity condition on the quartic potential $V_4\equiv\lambda_\text{h} h^4/4 +\lambda_\text{s} s^4/4+\lambda_\text{hs} h^2 s^2/2$ requires $\lambda_\text{h}>0, ~\lambda_\text{s}>0$, and 
\begin{equation}\label{pos}
 \lambda_\text{hs}\geq 0  \vee \left( \lambda_\text{hs}<0 \wedge \lambda^2_\text{hs}\leq \lambda_\text{h} \lambda_\text{s} \right)\,.
\end{equation}

$\bullet$ The coexistence condition is applied in a temperature interval where the first-order phase transition occurs; below a critical temperature and above a lower temperature both the false and true vacua must coexist. The lower temperature is usually at $T=0$, however in the scenario of {\it intermediate phase transition} that we investigate in this paper the lower temperature can lie at higher temperatures. 

$\bullet$ The true vacuum must remain the global minimum for all temperatures below the critical temperature and above a lower temperature. This lower temperature in the case of intermediate phase transition is non-zero.

For one- or two-step EWPT that we will discuss in sections \ref{secone-step} and \ref{sectwo-step}, whenever the singlet scalar takes zero VEV at $T=0$, the DM relic density can in principle be addressed by the model. For those scenarios in section \ref{secdm}, in addition to the EWPT conditions we will impose also the DM constraints including the restrictive bounds from direct detection experiments. We will also examine whether such DM scenarios are accessible in future colliders.

\section{One-Step Phase Transitions}\label{secone-step}
\subsection{Phase transition $(0,0)\to (v,0)$}\label{subsec00v0}
The vacuum $(0,0)$ is obviously an extremum of the thermal potential in Eq. (\ref{therpot}). In order for $(v,0)$ to be a stationary point we must have $v^2(T)=\mu^2_\text{h}(T)/\lambda_\text{h}$. Then the extremum $(v,0)$ will be also a local minimum if,
\begin{equation}\label{00v0}
 \mu^2_\text{h}(T)>0~~~\wedge~~~\frac{\lambda_\text{hs}}{\lambda_\text{h}}>\frac{\mu^2_\text{s}(T)}{\mu^2_\text{h}(T)}\,.
\end{equation}
The temperature-dependent mass matrix in the vacuum $(0,0)$ is diagonal and is given by $M\equiv \text{diag}(-\mu^2_\text{h}(T),-\mu^2_\text{s}(T))$. The minimum condition on $(0,0)$ then requires $\mu^2_\text{h}(T)<0$ and $\mu^2_\text{s}(T)<0$ which is in contrast with one of the conditions in Eq. (\ref{00v0}). 
Note that the two extrema $(0,0)$ and $(v,0)$ must coexist from a critical temperature $T^2_c=\mu^2_\text{h}/c_\text{h}$
down to $T=0$\footnote{This is a strong condition which is sufficient but not necessary. In general the false vacuum may disappear before reaching the lower limit of the phase transition temperature interval. For the simple $\mathbb{Z}_2$ symmetric potential in Eq. (\ref{therpot}) this is not the case.}. However as mentioned above, the minima $(0,0)$ and $(v,0)$ cannot coexist at the same time, therefore the phase transition cannot be a first-order one. Nevertheless, $(0,0)$ can be a maximum if $\mu^2_\text{h}(T)>0$ and $\mu^2_\text{s}(T)>0$ which is consistent with Eq. (\ref{00v0}). Hence, only a  second-order phase transition from $(0,0)$ to $(v,0)$ is plausible.\footnote{Note that the analytical perturbative investigation of the first-order electroweak phase transition in this paper is done using the leading order approximation at high temperature. In a non-perturebative computation in \cite{Gould:2019qek}, it is shown than the first-order phase transition $(0,0)\to (v,0)$ is possible for certain values of the parameters which is due to the inclusion of the next to leading order thermal contribution. In our analysis the CW contribution is {\it negligible} because we have assumed the high-temperature approximation. The full investigation to cover the low-temperature contributions is beyond our analytical approach.}
The conditions for a second-order phase transition are $\mu^2_\text{s}<0$, $\mu^2_\text{h}>0$, $c_\text{s}>0$, $\lambda_\text{s}>0$ and, 
\begin{equation}\label{00v02nd}
 \left( \lambda_\text{hs}\geq 0 \wedge \lambda_\text{h}>0 \right) \vee \left( \lambda_\text{hs}<0 \wedge \lambda_\text{h}\geq \frac{\lambda_\text{hs}\mu^2_\text{h}}{\mu^2_\text{s}} \right)\,.
\end{equation}

\subsection{Phase transition $(0,0)\to (v,w)$ }\label{subsec00vw}
  
As mentioned in the last section the point $(0,0)$ in the field configuration is clearly an extremum of the thermal effective potential. At least at the extreme of very hot universe, $T\to \infty$, the VEV solution $(0,0)$ must be a minimum (when the electroweak symmetry is unbroken). This imposes the conditions $c_\text{h}>0$ and $c_\text{s}>0$. On the other hand, the vacuum $(v,w)$ is an extremum solution if,
\begin{equation}\label{vevT}
 v^2(T)=\frac{\lambda_\text{hs}\mu^2_\text{s}(T)-\lambda_\text{s}\mu^2_\text{h}(T)}{\lambda_\text{hs}^2-\lambda_\text{h} \lambda_\text{s}}\hspace{2cm} w^2(T)=\frac{\lambda_\text{hs}\mu^2_\text{h}(T)-\lambda_\text{h}\mu^2_\text{s}(T)}{\lambda_\text{hs}^2-\lambda_\text{h} \lambda_\text{s}}\,.
\end{equation}
At zero temperature $v(T=0)\equiv v_{EW}=246$ GeV. Using this equality, the number of  free parameters can be reduced by one having $\mu^2_\text{h}= (\mu^2_\text{s}-\lambda_\text{hs}v_{EW}^2)\lambda_\text{hs}/\lambda_\text{s}+\lambda_\text{h} v_{EW}^2$.
 
The critical temperature at which the thermal potential has two degenerate minima is given by \footnote{Among two critical temperatures obtained from a generic effective potential, one is usually discarded when one imposes constraints such as the global minimum condition for the true vacuum, the coexistence condition for local minima in the first-order phase transition, the Higgs electroweak VEV at $T=0$, etc.},
\begin{equation}\label{Tcpm}
 T^2_{c\pm}=\frac{a \pm |b| \sqrt{\lambda}}{c}
\end{equation}
where 
\begin{equation}\label{a1a2bc}
\begin{split}
&\lambda=\lambda_\text{hs}^2 - \lambda_\text{h} \lambda_\text{s}\\
 &a=\left( c_\text{h} \lambda_\text{s}-c_\text{s} \lambda_\text{hs}  \right)\mu_\text{h}^2 + \left( c_\text{s} \lambda_\text{h} -c_\text{h} \lambda_\text{hs} \right) \mu_\text{s}^2 \\
 &b=c_\text{s} \mu_\text{h}^2 - c_\text{h} \mu_\text{s}^2\\
 &c=c_\text{s}^2 \lambda_\text{h} - 2 c_\text{h} c_\text{s} \lambda_\text{hs} + c_\text{h}^2 \lambda_\text{s}\,.
 \end{split}
\end{equation}
From Eq. (\ref{Tcpm}) we see that $\lambda\equiv \lambda^2_\text{hs} -\lambda_\text{h} \lambda_\text{s}\geq 0$ which according to Eq. (\ref{pos}) means $\lambda_\text{hs}\geq 0$.

Furthermore, Eq. (\ref{vevT}) must be positive for all temperatures including $T=0$. This condition alongside the positivity condition in Eq. (\ref{pos}) and $\lambda \geq 0$ implies $\mu^2_\text{s}>0$ and 
  \begin{equation}\label{posv0w0}
  \frac{\lambda_\text{h}  \mu_\text{s}^2}{\lambda_\text{hs}} <\mu_\text{h}^2 <\frac{\lambda_\text{hs} \mu_\text{s}^2 }{\lambda_\text{s}} \,.    
 \end{equation}

 The mass matrix for the potential in Eq. (\ref{3levelpot}) after the  symmetry breaking at $T=0$ is off-diagonal, therefore  there is a mixing between the Higgs scalar and the singlet scalar. Denoting the VEV of the scalars at $T=0$ by $(v_{EW},w_{EW})$ and rotating the scalar field configuration by an angle,
\begin{equation}
 \tan(\theta)= \frac{4v_{EW} w_{EW} \lambda_\text{hs}}{v_{EW}^2 (\lambda_\text{hs}-3 \lambda_\text{h})-w_{EW}^2(\lambda_\text{hs}-3\lambda_\text{s})+\mu^2_\text{h}-\mu^2_\text{s}}
\end{equation}
the mass matrix becomes diagonal. Using Eq. (\ref{vevT}) at $T=0$ for $(v_{EW},w_{EW})$ the mass eigenvalues are 
\begin{equation}\label{mass}
 m^2_\pm =  \frac{1}{\lambda} \left(a \pm \sqrt{4 \lambda b c +a^2}  \right)
\end{equation}
where
\begin{subequations}
\begin{align}
 &a=\lambda_\text{hs}\lambda_\text{s}\mu^2_\text{h}+\lambda_\text{h} \lambda_\text{hs}\mu^2_\text{s}-\lambda_\text{h}\lambda_\text{s}(\mu^2_\text{h}+\mu^2_\text{s})\\
 \begin{split}
 &b=\lambda_\text{hs}\mu^2_\text{h} -\lambda_\text{h}\mu^2_\text{s}\\
 &c=\lambda_\text{hs}\mu^2_\text{s} -\lambda_\text{s}\mu^2_\text{h} \,.
 \end{split}
 \end{align}
\end{subequations}

Both mass eigenvalues in Eq.(\ref{mass}) must be positive. However, the positivity of $m^2_-$ is in conflict with the condition in Eq. (\ref{posv0w0}). Therefore the critical temperature $T_{c-}$ is discarded by looking at the mass spectrum at zero temperature. Furthermore, one should require that the temperature-dependent mass for all $T\leq T_{c+}$ be positive. This is equivalent to have, 
\begin{equation}\label{minvwt}
 \frac{\mu^2_\text{h}(T)}{\mu^2_\text{s}(T)}<\frac{\lambda_\text{hs}}{\lambda_\text{s}} ~\wedge~  \frac{\mu^2_\text{h}(T)}{\mu^2_\text{s}(T)}<\frac{\lambda_\text{h}}{\lambda_\text{hs}}\,.
\end{equation}
The condition $m^2_+>0$ is now in conflict with Eq. (\ref{minvwt}). Therefore, also the transition from $(0,0)$ to $(v,w)$ via a single step in the singlet scalar model with $\mathbb{Z}_2$ symmetry cannot be of first-order type.  

A quicker way to show that the phase transition from $(0,0)$ to $(v,w)$ is not possible, neither through  first- nor second-order types, is as follows:
First, the VEV's $v$ and $w$ must be real and positive for all temperatures; let us take $T=0$ here. This requirement together with the positivity condition in Eq. (\ref{pos}) and  $\lambda\geq 0$ from the critical temperature in Eq. (\ref{Tcpm}) leads to $\lambda_\text{s}>0, \lambda_\text{hs}>0, \mu^2_\text{s}>0$, and
\begin{equation}\label{vwpos}
0<\lambda_\text{h}<\frac{\lambda^2_\text{hs}}{\lambda_\text{s}} ~\wedge~ \frac{\lambda_\text{h}\mu^2_\text{s}}{\lambda_\text{hs}} < \mu^2_\text{h} <  \frac{\lambda_\text{hs}\mu^2_\text{s}}{\lambda_\text{s}}\,.
\end{equation}
Then the minimum condition for $(v,w)$ at $T=0$ --i.e. for $(v_{EW},w_{EW})$-- using Eq. (\ref{vevT}) is, 
\begin{equation}
 \lambda_\text{s} \mu^2_\text{h} < \lambda_\text{hs} \mu^2_\text{s} ~\wedge~ \lambda_\text{hs}\mu^2_\text{h} < \lambda_\text{h} \mu^2_\text{s}
\end{equation}
which is immediately in conflict with Eq. (\ref{vwpos}), hence no first or second-order phase transition becomes possible for this channel.
\footnote{Note that in the quick proof we have only used the conditions at $T=0$. In the longer proof we have used the temperature-dependent minimum condition in Eq. (\ref{minvwt}).}

\section{Two-Step Phase Transitions}\label{sectwo-step}
Here we consider the electroweak phase transitions that an intermediate step intervene the process to reach us to the vacuum at $T=0$. In the last section it was argued that a one-step first-order phase transition is never possible for $\mathbb{Z}_2$ symmetric singlet scalar model. In the following sections we examine the first- or second-order phase transitions in either steps. A scenario of the two-step phase transition is considered in \cite{Kurup:2017dzf}.
\subsection{Phase Transition $(0,0)\stackrel{\text{1st or 2nd}}{\longrightarrow}(0,w)\stackrel{\text{1st}}{\to}(v,0)$}\label{subsec000wv0}
We examine whether a two-step phase transition is possible with a first-order phase transition in the second step from $(0,w)$ to $(v,0)$. For the moment we relax the type of the phase transition in the first step.

The temperature-dependent VEVs which are the extrema of the thermal effective potential in Eq. (\ref{therpot}) are,
\begin{equation}
 w^2(T)=\frac{\mu^2_\text{s}-c_\text{s} T^2}{\lambda_\text{s}},~~~~~v^2(T)=\frac{\mu^2_\text{h}-c_\text{h} T^2}{\lambda_\text{h}}\,.
\end{equation}

To begin with, we determine the critical temperature at which each step gets triggered. The first and second critical temperature read, 
\begin{equation}
 T^2_c=\frac{\mu^2_\text{s}}{c_\text{s}},~~~~T'^2_{c-}=\frac{\mu^2_\text{s}\sqrt{\lambda_\text{h}}-\mu^2_\text{h}\sqrt{\lambda_\text{s}}}{c_\text{s}\sqrt{\lambda_\text{h}} -c_\text{h}\sqrt{\lambda_\text{s}}},~~~~T'^2_{c+}=\frac{\mu^2_\text{s}\sqrt{\lambda_\text{h}}+\mu^2_\text{h}\sqrt{\lambda_\text{s}}}{c_\text{s}\sqrt{\lambda_\text{h}}+c_\text{h}\sqrt{\lambda_\text{s}}}, 
\end{equation}
where from Eq. (\ref{pos}), $\lambda_\text{h}$ and $\lambda_\text{s}$ are both positive. At this stage there are two critical temperatures for the second step. The positivity of $T^2_c$ together with Eq. (\ref{chcs}) immediately results in $\mu^2_\text{s}>0$. Moreover, $T^2_c>T'^2_{c\pm}$ leads to, 
\begin{subequations}\label{T1>T2}
\begin{align}\label{T1>T2-}
&T^2_c>T'^2_{c-}:~~~ \left( \frac{\mu^4_\text{h}}{\mu^4_\text{s}}<\frac{\lambda_\text{h}}{\lambda_\text{s}}~\wedge~ \frac{\mu^2_\text{h}}{\mu^2_\text{s}}>\frac{c_\text{h}}{c_\text{s}}\right)~\vee~ \left( \frac{\mu^2_\text{h}}{\mu^2_\text{s}}<\frac{c_\text{h}}{c_\text{s}}~\wedge~ \frac{\lambda_\text{h}}{\lambda_\text{s}}<\frac{\mu^4_\text{h}}{\mu^4_\text{s}} \right) \\\label{T1>T2+}
&T^2_c>T'^2_{c+}:~~~ \left( \frac{\mu^2_\text{h}}{\mu^2_\text{s}}<\frac{c_\text{h}}{c_\text{s}} ~\wedge~ \mu^2_\text{h} > 0 \right)~\vee~ \left( \mu^2_\text{h} \leq 0 ~\wedge~ \frac{\lambda_\text{h}}{\lambda_\text{s}}>\frac{\mu^4_\text{h}}{\mu^4_\text{s}} \right)\,.
\end{align}
\end{subequations}
We have assumed that the stationary point $(0,0)$ is a global minimum at high temperature from which the condition in Eq.(\ref{chcs}) must be satisfied. At temperature $T_c$ and afterwards the vacuum solution $(0,0)$ is no longer a global minimum, instead the extremum $(0,w)$ becomes the deeper minimum down to a second critical temperature $T'_{c\pm}$. For all temperatures below $T'_{c\pm}$ both minima $(0,w)$ and $(v,0)$ must coexist. In addition, the minimum $(v,0)$ should remain the deeper one for all $T<T'_{c\pm}$. 

In order for the vacuum solutions $(0,w)$ and $(v,0)$ to be coexisting minima for $T<T'_{c\pm}$ the following conditions respectively must be satisfied at the same time,

\begin{subequations}
\begin{align}
\mu^2_\text{s}(T)>0 ~\wedge~  \frac{\lambda_\text{hs}}{\lambda_\text{s}} \mu^2_\text{s}(T)>\mu^2_\text{h}(T) \label{cminw0}\\
 \mu^2_\text{h}(T)>0 ~\wedge~  \frac{\lambda_\text{hs}}{\lambda_\text{h}} \mu^2_\text{h}(T)>\mu^2_\text{s}(T) \,. \label{cminv0}
 \end{align}
\end{subequations}
Another way to ensure that the extremum $(v,0)$ remains always a minimum for $0 \leq T\leq T_{c\pm}$ is to impose that the mass matrix possesses positive eigenvalues in this temperature interval, 
\begin{subequations}\label{masv0}
\begin{align}
& m_h(T)\equiv 2(\mu^2_\text{h}-c_\text{h} T^2)> 0 ~~~~~&&\forall T\leq T'_{c\pm}\\
& m_s(T)\equiv -(\mu^2_\text{s}-c_\text{s} T^2)+\frac{\lambda_\text{hs}}{\lambda_\text{s}}(\mu^2_\text{h}-c_\text{h} T^2)> 0~~~~~&&\forall T\leq T'_{c\pm}
 \end{align}
\end{subequations}
which are equivalent to Eq. (\ref{cminv0}).
Eq. (\ref{T1>T2+}) is not consistent with the conditions in Eq. (\ref{cminv0}) or Eqs. (\ref{masv0}). Therefore the correct critical temperature is $T'_{c-}$. Combining Eqs. (\ref{chcs}), (\ref{T1>T2-}), (\ref{cminw0}) and (\ref{cminv0}) would lead to the coexistence condition for the two minima below the critical temperature,
\begin{equation}\label{coex}
 \left( \frac{\lambda_\text{hs}}{\lambda_\text{s}}>\frac{\mu^2_\text{h}}{\mu^2_\text{s}}~\wedge~ \frac{c_\text{h}}{c_\text{s}}>\frac{\mu^2_\text{h}}{\mu^2_\text{s}}~\wedge~\frac{\lambda_\text{h}}{\lambda_\text{s}}<\frac{\mu^4_\text{h}}{\mu^4_\text{s}} \right)~\vee~  \left( \frac{c_\text{h}}{c_\text{s}}<\frac{\mu^2_\text{h}}{\mu^2_\text{s}}~\wedge~\lambda_\text{s}\frac{\mu^4_\text{h}}{\mu^4_\text{s}}< \lambda_\text{h} \leq  \lambda_\text{h}\frac{\mu^2_\text{h}}{\mu^2_\text{s}} \right)\,. \\
\end{equation}
We should also impose that the minimum $(v,0)$ be deeper than the minimum $(w,0)$ in all temperatures below $T'_{c-}$ i.e. , 
\begin{equation}\label{globT}
 \frac{\mu^4_\text{h}(T)}{\lambda_\text{h}}>\frac{\mu^4_\text{s}(T)}{\lambda_\text{s}} ~~~~~~\forall T<T'_{c-} \,.
\end{equation}
The above $T$-dependent requirement when taking into account also Eqs. (\ref{chcs}) and (\ref{T1>T2-}) gives rise to, 
\begin{equation}\label{glob}
\begin{split}
 &\left( -\lambda_\text{s} \frac{\mu^2_\text{h}}{\mu^2_\text{s}}<\lambda_\text{hs} <0~\wedge~ \frac{c_\text{h}}{c_\text{s}}>\frac{\mu^2_\text{h}}{\mu^2_\text{s}}~\wedge~  \frac{\lambda^2_\text{hs}}{\lambda_\text{s}} \leq \lambda_\text{h}<\lambda_\text{s} \frac{\mu^4_\text{h}}{\mu^4_\text{s}} \right) \\
 &\hspace{2cm}\vee \left(\lambda_\text{hs}\geq 0 ~\wedge~ \frac{c_\text{h}}{c_\text{s}}>\frac{\mu^2_\text{h}}{\mu^2_\text{s}}~\wedge~ 0<\lambda_\text{h}<\lambda_\text{s} \frac{\mu^4_\text{h}}{\mu^4_\text{s}} \right)\,. \\
 \end{split}
\end{equation}
All in all, the combination of the coexistence condition in Eq. (\ref{coex}) and the global minimum condition in Eq. (\ref{glob}), the necessary condition for a first-order phase transition in the second step  to happen in the aforementioned two-step channel, is given by $\lambda_\text{h}>0,~\lambda_\text{s}>0,~ c_\text{h}>0 ,~ c_\text{s}>0,~\mu^2_\text{h}>0 ,~\mu^2_\text{s}>0$, and 
\begin{equation}\label{000wv0}
 \frac{\lambda_\text{h}}{\lambda_\text{s}}<\frac{\mu^4_\text{h}}{\mu^4_\text{s}}~\wedge~ \frac{c_\text{h}}{c_\text{s}}>\frac{\mu^2_\text{h}}{\mu^2_\text{s}}~\wedge~    \frac{\lambda_\text{hs}}{\lambda_\text{s}} \geq \frac{\mu^2_\text{h}}{\mu^2_\text{s}} \,.
\end{equation} 
Let us now investigate whether also the first step i.e. $(0,0)$ to $(0,w)$ can be a first- or second-order phase transition? In addition to the final result in Eq. (\ref{000wv0}), we should add the coexistence condition of the minima at $(0,0)$ and $(0,w)$ in the temperature interval   $T_c<T<T'_{c-}$. The coexistence condition that supports the existence of a barrier for the first-order phase transition, together with positivity of the quartic potential and the sequence of two phase transitions being $T_c>T'_{c-}$, give rise to $\lambda_\text{s}>0$, $\mu^2_\text{h}>0$, $\mu^2_\text{s}>0$,  $c_\text{s}>0$ and,

if $\lambda_\text{hs}<-\lambda_\text{s} \mu^2_\text{h} /\mu^2_\text{s}$, 
\begin{equation}\label{000w}
 \lambda_\text{h}\geq \frac{\lambda^2_\text{hs}}{\lambda_\text{s}} ~\wedge~ 0 <c_\text{h}<\frac{c_\text{s} \mu^2_\text{h}}{\mu^2_\text{s}}
\end{equation}

if $-\lambda_\text{s} \mu^2_\text{h} /\mu^2_\text{s}<\lambda_\text{hs}\leq 0$, 
\begin{equation}
 \left( \frac{\lambda^2_\text{hs} }{\lambda_\text{s}}\leq\lambda_\text{h}< \frac{\lambda_\text{s} \mu^4_\text{h}}{\mu^4_\text{s}} ~\wedge ~ c_\text{h}>\frac{c_\text{s} \mu^2_\text{h}}{\mu^2_\text{s}} \right) ~\vee~  \left( \lambda_\text{h}> \frac{\lambda_\text{s} \mu^4_\text{h}}{\mu^4_\text{s}} ~\wedge~ 0<c_\text{h}<\frac{c_\text{s} \mu^2_\text{h}}{\mu^2_\text{s}}\right)
\end{equation}

if $\lambda_\text{hs}>0$, 
\begin{equation}
\left( 0 \leq  \lambda_\text{h}< \frac{\lambda_\text{s} \mu^4_\text{h}}{\mu^4_\text{s}} ~\wedge~ c_\text{h}>\frac{c_\text{s} \mu^2_\text{h}}{\mu^2_\text{s}} \right) ~\vee~ \left( \lambda_\text{h}> \frac{\lambda_\text{s} \mu^4_\text{h}}{\mu^4_\text{s}} ~\wedge~ 0<c_\text{h}<\frac{c_\text{s} \mu^2_\text{h}}{\mu^2_\text{s}}\right) \,. 
\end{equation}
The overlap of the conditions in Eq. (\ref{000w}) for the first-order phase transition in the first step, and the conditions in Eq. (\ref{000wv0}) for the first-order phase transition in the second step will be again Eq. (\ref{000wv0}). A similar analysis shows that a second-order phase transition in the first step is also possible if conditions in Eq. (\ref{000wv0}) is fulfilled. 

If we relax the first-order phase transition condition in the second step and instead require only the minimum condition for $(v,0)$ at $T=0$ which is $\lambda_\text{h}>0$ and,
\begin{equation}
 \mu^2_\text{h}>0 ~\wedge~  \lambda_\text{hs}<\frac{\lambda_\text{h} \mu^2_\text{s}}{\mu^2_\text{h}}
\end{equation}
then the condition for a first-order phase transition to happen only in the first step will be $\mu^2_\text{h}>0$, $\mu^2_\text{s}>0$, $\lambda_\text{s}>0$, $c_\text{s}>0$ and, 

if $0<\lambda_\text{hs} \leq \lambda_\text{s}\mu^2_\text{h}/ \mu^2_\text{s}$: 
\begin{equation}
 0<\lambda_\text{h}<\frac{\lambda_\text{hs}\mu^2_\text{h}}{\mu^2_\text{s}} ~\wedge~ c_\text{h}>\frac{c_\text{s}\mu^2_\text{h}}{\mu^2_\text{s}}
\end{equation}

if $\lambda_\text{hs} > \lambda_\text{s}\mu^2_\text{h}/ \mu^2_\text{s}$:
\begin{equation}
 \left( c_\text{h}>\frac{c_\text{s}\mu^2_\text{h}}{\mu^2_\text{s}} ~\wedge~ 0<\lambda_\text{h}<\frac{\lambda_\text{s}\mu^4_\text{h}}{\mu^4_\text{s}} \right) \vee \left(0<c_\text{h}<\frac{c_\text{s}\mu^2_\text{h}}{\mu^2_\text{s}} ~\wedge~ \frac{\lambda_\text{s}\mu^4_\text{h}}{\mu^4_\text{s}}<\lambda_\text{h}<\frac{\lambda_\text{hs}\mu^2_\text{h}}{\mu^2_\text{s}}\right)\,.
\end{equation}

\subsection{Phase Transition $(0,0)\to(0,w)\stackrel{\text{2nd}}{\to}(v',w')$}\label{subsec000wvw}
Here we discuss a two-step scenario with the assumption that the second step i.e. $(0,w)\to(v',w')$ is a first-order phase transition. We do not impose any conditions for the first step $(0,0)\to(0,w)$ for the moment which means it can be a first-order, second-order or crossover phase transition. 
Here the singlet scalar takes a non-zero VEV also in the second step, so the $\mathbb{Z}_2$ symmetry remains broken down to low temperatures. 

The extremum solution in the last step is given by,
\begin{equation}
 v'^2(T)=\frac{\lambda_\text{hs}\mu^2_\text{s}(T)-\lambda_\text{s}\mu^2_\text{h}(T)}{\lambda_\text{hs}^2-\lambda_\text{h} \lambda_\text{s}}\hspace{2cm} w^2(T)=\frac{\lambda_\text{hs}\mu^2_\text{h}(T)-\lambda_\text{h}\mu^2_\text{s}(T)}{\lambda_\text{hs}^2-\lambda_\text{h} \lambda_\text{s}}
\end{equation}
while in the first step the VEV is as follows, 
\begin{equation}
 w^2(T)=\frac{\mu^2_\text{s}(T)}{\lambda_\text{s}}\,.
\end{equation}
The first and second critical temperature for this phase transition channel read, 
\begin{equation}
 T_c=\frac{\mu^2_\text{s}}{c_\text{s}}\hspace{4cm}
 T'_c=\frac{\lambda_\text{s}\mu^2_\text{h}-\lambda_\text{hs}\mu^2_\text{s}}{c_\text{h}\lambda_\text{s}-\lambda_\text{hs}c_\text{s}}
\end{equation}
where $T_c>0$ immediately gives $\mu^2_\text{s}>0$. Above the temperature $T_c$ the vacuum solution $(0,0)$ is a global minimum, which requires Eq. (\ref{chcs}), while below $T_c$ it becomes a maximum. The first-order phase transition in the second step triggers at a lower temperature $T'_c$, i.e. $0<T'_c<T_c$. Both solutions $(0,w)$ and $(v',w')$ must coexist in the temperature interval $0 \leq T < T'_c$. In addition $(v',w')$ must remain the global minimum for all temperature below $T'_c$ down to $T=0$. 
The condition $0<T'_c<T_c$ imposes the following bounds on the parameters; $\mu^2_\text{s}>0$ and,
\begin{equation}\label{t1>t2}
 \begin{split}
  &\text{if}~ \lambda_\text{hs}\geq 0:  \left(c_\text{h}<\frac{c_\text{s}\lambda_\text{hs}}{\lambda_\text{s}}\wedge \frac{c_\text{s}\mu^2_\text{s}}{c_\text{s}}< \mu^2_\text{h} <\frac{\lambda_\text{hs}\mu^2_\text{s}}{\lambda_\text{s}}\right) \\
  &~~~~~~~~~~~~~\vee \left( c_\text{h}>\frac{c_\text{s}\lambda_\text{hs}}{\lambda_\text{s}}\wedge \frac{\lambda_\text{hs}\mu^2_\text{s}}{\lambda_\text{s}}< \mu^2_\text{h} <\frac{c_\text{h}\mu^2_\text{s}}{c_\text{s}} \right) \\
 &\text{if}~ \lambda_\text{hs}<0: \left( \frac{\lambda_\text{hs}\mu^2_\text{s}}{\lambda_\text{s}}< \mu^2_\text{h} <\frac{c_\text{h}\mu^2_\text{s}}{c_\text{s}} \wedge \lambda_\text{h}\geq \frac{\lambda^2_\text{hs}}{\lambda_\text{s}} \right).
   \end{split}
\end{equation}
On the one hand, the minimum condition on the extremum solution $(0,w)$ for $0 \leq T< T'_c$, taking into account Eq. (\ref{t1>t2}) leads to $\lambda_\text{hs}>0 ,~ \mu^2_\text{s}>0$ and
\begin{equation}\label{m0w}
 0<c_\text{h}<\frac{c_\text{s}\lambda_\text{hs}}{\lambda_\text{s}}\wedge \frac{c_\text{h}\mu^2_\text{s}}{c_\text{s}} < \mu^2_\text{h} < \frac{\lambda_\text{hs}\mu^2_\text{s}}{\lambda_\text{s}}\,.
\end{equation}
On the other hand, considering Eq. (\ref{t1>t2}), the vacuum solution $(v',w')$ will be a minimum for all temperature $T\leq T'_c$ provided that $\mu^2_\text{s}>0$ and,
\begin{equation}\label{mvw}
 \begin{split}
  &\text{if}~\mu^2_\text{h}\leq 0: \lambda_\text{hs}<\frac{\lambda_\text{s}\mu^2_\text{h}}{\mu^2_\text{s}}\wedge \lambda_\text{h}>\frac{\lambda^2_\text{hs}}{\lambda_\text{s}}\\
  & \text{if}~\mu^2_\text{h}>0: 
  \left(\lambda_\text{hs}\leq 0 \wedge c_\text{h}>\frac{c_\text{s}\mu^2_\text{h}}{\mu^2_\text{s}} \wedge \lambda_\text{h}>\frac{\lambda^2_\text{hs}}{\lambda_\text{s}}  \right) , \\
&~~~~~~~~~~~~~~  \vee  \left(0 <\lambda_\text{hs}< \frac{\lambda_\text{s}\mu^2_\text{h}}{\mu^2_\text{s}} \wedge c_\text{h}>\frac{c_\text{s}\mu^2_\text{h}}{\mu^2_\text{s}} \wedge \lambda_\text{s} \geq \frac{\lambda_\text{h} \mu^2_\text{h}}{\mu^2_\text{s}}
  \right).
 \end{split}
\end{equation}
It is not difficult to show that Eqs. (\ref{m0w}) and (\ref{mvw}) cannot be satisfied at the same time so that the minima $(0,w)$ and $(v',w')$ do not coexist in the interval $T\leq T'_c$. Therefore, regardless of the type of the phase transition in the first step, a first-order phase transition in the second step is not possible. 

Although the phase transition in the second step cannot be first-order but it can be of second-order type provided that $\mu^2_\text{s}>0$, $\lambda_\text{s}>0$,  $c_\text{s}>0$ and, 

if $\mu^2_\text{h}>0$:
\begin{equation}
\begin{split}
 &\left( 0<\lambda_\text{h}<\frac{\lambda_\text{s}\mu^4_\text{h}}{\mu^4_\text{s}} ~\wedge~ 
 c_\text{h} > \frac{c_\text{s} \mu^2_\text{h}}{\mu^2_\text{s}} ~\wedge~ -\sqrt{\lambda_\text{h} \lambda_\text{s}}<\lambda_\text{hs}\leq \frac{  \lambda_\text{h} \mu^2_\text{s}}{\mu^2_\text{h}}  \right) ~\vee~ \\
 &\left( \lambda_\text{h} \geq \frac{\lambda_\text{s}\mu^4_\text{h}}{\mu^4_\text{s}} ~\wedge~ c_\text{h} > \frac{c_\text{s} \mu^2_\text{h}}{\mu^2_\text{s}} ~\wedge~ -\sqrt{\lambda_\text{h}\lambda_\text{s}}<\lambda_\text{hs}< \frac{  \lambda_\text{s} \mu^2_\text{h}}{\mu^2_\text{s}}  \right)\,,
 \end{split}
\end{equation}

if  $\mu^2_\text{h} \leq 0$:
\begin{equation}
  \lambda_\text{h} >\frac{\lambda_\text{s}\mu^4_\text{h}}{\mu^4_\text{s}}  ~\wedge~ -\sqrt{\lambda_\text{h}\lambda_\text{s}}<\lambda_\text{hs}< \frac{  \lambda_\text{s} \mu^2_\text{h}}{\mu^2_\text{s}} \,.
\end{equation}
However, the first step in this scenario i.e. $(0,0)\to (0,w)$ is neither a first- nor a second-order phase transition. 

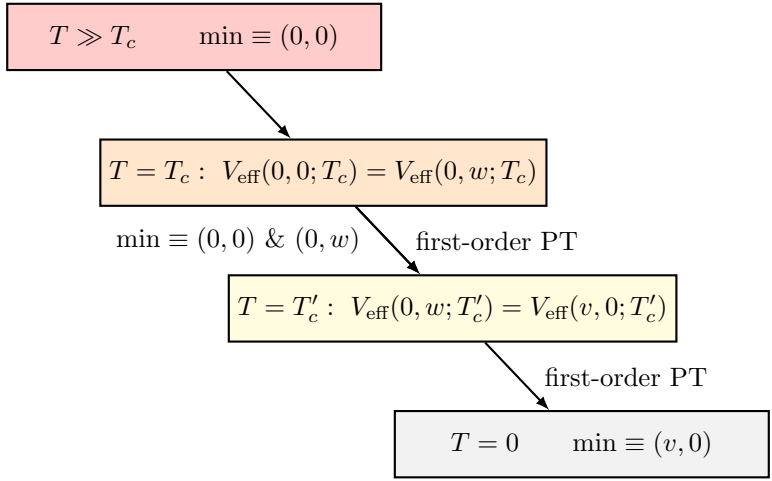
\begin{figure}\footnotesize
\tikzstyle{block} = [draw,thick,text centered, minimum height=2.5em,minimum width=14em]
    \centering\begin{tikzpicture}
    \node [block,fill=red!20] at (0,0) (1) {$T\gg T_c~~~~~~~\text{min}\equiv (0,0)$};
    \node [block,fill=orange!20] at (1.7,-1.8) (2) {$T=T_{c}:~ V_\text{eff}(0,0;T_{c})=V_\text{eff}(0,w;T_{c})$};
    \node [block,fill=yellow!15] at (3.4,-3.6) (3) { $T=T'_{c}:~ V_\text{eff}(0,w;T'_{c})=V_\text{eff}(v,0;T'_{c})$};
    \node [block,fill=gray!10] at (5.1,-5.4) (4) { $T=0~~~~~~\text{min}\equiv(v,0)$};
  (1) \draw [-latex, thick] (1) -- node [right] {$~~$} (2) ;
  \draw [-latex, thick] (2) --  node [left] {$~~\text{min}\equiv(0,0) ~\&~ (0,w)~~$} (3) ;
 \draw [-latex, thick] (2) --  node [right] {$~~\text{first-order PT}$} (3) ;
  \draw [-latex, thick] (3) --  node [right] {$~~\text{first-order PT}$} (4) ;
    \end{tikzpicture} 
    \caption{A first-order phase transition may trigger in the hot early universe with temperature $T_c$ and end in an intermediate lower temperature $T'_c$ .}\label{intephase}
    \end{figure}
\subsection{Phase Transition $(0,0)\to(v,w)\to(v',0)$}\label{subsec00vwv0}
Here we examine if this two-step phase transition can undergo a first-order phase transition which starts and ends in an intermediate temperature interval; i.e. if the vacuum $(0,0)$ can undergo a phase transition to $(v,w)$ at high temperature and then through a second phase transition to $(v',0)$ down to zero temperature. This type of first-order phase transition has been depicted in Fig. \ref{intephase} for the scenario in section \ref{subsec000wv0}.
The first and second critical temperatures at first and second steps read respectively,
\begin{subequations}
\begin{align}
\label{tcpm}
&T_{c\pm}=\frac{(c_\text{h}\lambda_\text{s}-c_\text{s}\lambda_\text{hs})\mu^2_\text{h}+(c_\text{s}\lambda_\text{h}-c_\text{h}\lambda_\text{hs})\mu^2_\text{s}\pm |c_\text{s}\mu^2_\text{h}-c_\text{h}\mu^2_\text{s}|\sqrt{\lambda^2_\text{hs}-\lambda_\text{h}\lambda_\text{s}}}
{c^2_\text{s}\lambda_\text{h}-2c_\text{h}c_\text{s}\lambda_\text{hs}+c^2_\text{h}\lambda_\text{s}} \\
&T'_c=\frac{\lambda_\text{hs}\mu^2_\text{h}-\lambda_\text{h}\mu^2_\text{s}}{c_\text{h}\lambda_\text{hs}-c_\text{s}\lambda_\text{hs}}\,.
\end{align}
\end{subequations}
From Eq. (\ref{tcpm}) $\lambda^2_\text{hs}>\lambda_\text{h}\lambda_\text{s}$ that together with the positivity condition in Eq. (\ref{pos}) it means $\lambda_\text{hs}\geq 0$. Furthermore, the sequence of the phase transitions to be first $(0,0)\to (v,w)$ and then $(v,w)\to(v',0)$ requires that $T'_c<T_{c\pm}$. 
A first-order phase transition should occur in the first step if both $(0,0)$ and $(v,w)$ minima exist for all temperature $T'_c<T<T_{c\pm}$. In other words the conditions, 
\begin{equation}\label{cc1}
 \mu^2_\text{h}(T)<0 ~\wedge~ \mu^2_\text{s}(T)<0
\end{equation}
for the solution $(0,0)$, and 
\begin{equation}\label{cc2}
 \lambda_\text{s}\mu^2_\text{h}(T)<\lambda_\text{hs}\mu^2_\text{s}(T) ~\wedge~ \lambda_\text{hs}\mu^2_\text{h}(T)<\lambda_\text{h}\mu^2_\text{s}(T) 
\end{equation}
for the solution $(v,w)$ must hold for all temperatures $T'_c<T<T_{c\pm}$. However, the minimum conditions for $(0,0)$ in Eq. (\ref{cc1}) and for $(v,w)$ in Eq. (\ref{cc2}) are not consistence with $T'_c<T_{c-}$. The minimum condition on  the solution $(0,0)$ for all temperature in the interval $T'_c<T<T_{c+}$ is given by, 
\begin{equation}\label{00}
\begin{split}
  &\frac{2c_\text{h}c_\text{s}\lambda_\text{hs}-c^2_\text{s}\lambda_\text{h}}{c^2_\text{h}}<\lambda_\text{s}<\frac{\lambda^2_\text{hs}}{\lambda_\text{h}} \wedge c_\text{h}>\frac{c_\text{s}\lambda_\text{h}}{\lambda_\text{hs}} \\
  &\hspace{2cm}\wedge  
   \left( (\mu^2_\text{s} \leq 0 \wedge \mu^2_\text{h}>0 
  ) \vee ( \mu^2_\text{s}>0 \wedge \mu^2_\text{h}>\frac{c_\text{h} \mu^2_\text{s}}{c_\text{s}}) \right) \,.
  \end{split}
\end{equation}
The minimum condition on $(v,w)$ for $T'_c<T<T_{c+}$ is given by $\lambda_\text{hs}>0$, $\lambda_\text{s}>0$, $\mu^2_\text{s}>0$, $\lambda_\text{h} \mu^2_\text{s}/\lambda_\text{hs}<\mu^2_\text{h}<c_\text{h} \mu^2_\text{s}/c_\text{s}$ and, 

if $0<c_\text{h}<c_\text{s} \lambda_\text{hs}/\lambda_\text{s}$:
\begin{equation}
 0<\lambda_\text{h}<\frac{c_\text{h} \lambda_\text{hs}}{c_\text{s}}
\end{equation}

if $c_\text{h}\geq 2 c_\text{s} \lambda_\text{hs}/\lambda_\text{s}$:

\begin{equation}
 0<\lambda_\text{h}<\frac{ \lambda^2_\text{hs}}{\lambda_\text{s}}
\end{equation}

if $c_\text{s} \lambda_\text{hs}/\lambda_\text{s}<c_\text{h}<2 c_\text{s} \lambda_\text{hs}/\lambda_\text{s}$:

\begin{equation}
 0<\lambda_\text{h}<\frac{ c_\text{h}\left( 2 c_\text{s} \lambda_\text{hs}- c_\text{h} \lambda_\text{s}\right) }{c^2_\text{s}} ~\vee~  \frac{ c_\text{h}\left( 2 c_\text{s} \lambda_\text{hs}- c_\text{h} \lambda_\text{s}\right) }{c^2_\text{s}}<\lambda_\text{h}<\frac{ \lambda^2_\text{hs}}{\lambda_\text{s}}
\end{equation}
which is in conflict with Eq. (\ref{00}). In other words, the minima $(0,0)$ and $(v,w)$ do not coexist in the temperature interval $T'_c<T<T_{c+}$. Therefore an intermediate temperature first-order phase transition in the first step is not possible. Similar analysis shows that even a second-order phase transition cannot occur in the first step. 

Next we answer the question whether in the second step, a first- or second-order phase transition is achievable. 
To have a barrier between minima in the second step, the two minima $(v,w)$ and $(v',0)$ must be both local minima for all values of temperature within $0<T<T'_c$. This is possible if $\mu^2_\text{h}>0$, $\mu^2_\text{s}>0$, $\lambda_\text{s}>0$, $c_\text{s}>0$ and, 
\begin{equation}\label{cominvwv0}
 0<\lambda_\text{h}< \frac{\lambda_\text{s}\mu^4_\text{h}}{\mu^4_\text{s}} ~\wedge~ c_\text{h}=\frac{c_\text{s}\mu^2_\text{h}}{\mu^2_\text{s}} ~\wedge~ \lambda_\text{hs}=\frac{c_\text{h}\lambda_\text{s}}{c_\text{s}}
\end{equation}
in which there are two matching conditions. The condition when no barrier appears between the minima is $\mu^2_\text{h}>0$, $\mu^2_\text{s}>0$, $\lambda_\text{s}>0$, $c_\text{s}>0$ and, 
\begin{equation}\label{comaxvwv0}
 0<\lambda_\text{h}< \frac{\lambda_\text{hs}\mu^2_\text{h}}{\mu^2_\text{s}} ~\wedge~ c_\text{h}=\frac{c_\text{s}\mu^2_\text{h}}{\mu^2_\text{s}} ~\wedge~ \lambda_\text{hs}=\frac{\lambda_\text{s}\mu^2_\text{h}}{\mu^2_\text{s}}\,.
\end{equation}
Furthermore, $(v',0)$ in the second step must be the deepest minimum for all temperature $0<T<T'_c$, i.e. $\Delta V\equiv V(v,w;T)-V(v',0;T)>0$, for all $0<T<T'_c$\,, 
\begin{equation}
 \frac{\left[\lambda_\text{hs}\mu^2_\text{h}(T)-\lambda_\text{h}\mu^2_\text{s}(T)\right]^2}{4\lambda_\text{h}\left(\lambda_\text{hs}-\lambda_\text{h}\lambda_\text{s}\right)}>0
\end{equation}
which implies $\lambda_\text{hs}>\lambda_\text{h}\lambda_\text{s}$. This condition is in conflict with Eqs. (\ref{cominvwv0}) and (\ref{comaxvwv0}), therefore also in the second step no phase transition of first- or second-order type is plausible.

\subsection{Phase Transition $(0,0)\to(v,0)\to(v',w')$}\label{00v0vw}
Is a first-order electroweak transition ever allowed in a two-step phase transition and during an intermediate temperature interval in the real singlet scalar model with $\mathbb{Z}_2$ symmetry? We finish this section by answering this question in the last scenario of the two-step phase transition scenarios, i.e. a phase transition from $(0,0)$ to $(v,0)$ followed by $(v,0)$ to $(v',w')$. The first and second critical temperatures for this scenario are, 
\begin{equation}
T_c=\frac{\mu^2_\text{h}}{\lambda_\text{h}}\,,\hspace{2cm}
T'_c=\frac{\lambda_\text{hs}\mu^2_\text{h}-\lambda_\text{h}\mu^2_\text{s}}{c_\text{h}\lambda_\text{hs}-c_\text{s}\lambda_\text{hs}}\,.
\end{equation}
The requirement $0<T'_c<T_c$ leads to $\mu^2_\text{h}>0$ and,
\begin{equation}\label{T2<T1}
 \left(\lambda_\text{hs} > \frac{\lambda_\text{h}\mu^2_\text{s}}{\mu^2_\text{h}} ~\wedge~ c_\text{s} < \frac{c_\text{h}\mu^2_\text{s}}{\mu^2_\text{h}} \right) ~\vee~ \left( c_\text{s} > \frac{c_\text{h}\mu^2_\text{s}}{\mu^2_\text{h}} ~\wedge~ \lambda_\text{hs} < \frac{\lambda_\text{h}\mu^2_\text{s}}{\mu^2_\text{h}}\right) \,.
\end{equation}

At $T=0$ the scalars masses must be real and positive. The mass matrix is given by Eq. (\ref{hessian}) with $(h,s)\equiv (v',w')$. The mass eigenvalues after diagonalizing the mass matrix are already obtained in Eq. (\ref{mass}). Another set of the conditions therefore come from $m^2_{\pm}>0$ which considering $\mu^2_\text{s}>0$ and $\lambda_\text{h} \lambda_\text{s} \geq \lambda^2_\text{hs}$ from the positivity becomes,
\begin{equation}\label{m2>0}
\left(\lambda_\text{hs}\leq 0 ~\wedge~  \mu^2_\text{s} > \frac{\lambda_\text{hs}\mu^2_\text{h}}{\lambda_\text{h}} \right) ~\vee~ \left(\lambda_\text{hs}> 0  ~\wedge~\frac{\lambda_\text{hs}\mu^2_\text{h}}{\lambda_\text{h}} <\mu^2_\text{s}< \frac{\lambda_\text{s}\mu^2_\text{h}}{\lambda_\text{hs}}\right) \,.
\end{equation}
The minimum conditions for the vacuum solutions $(0,0)$ and $(v,0)$ taking into account Eqs. (\ref{chcs}) and (\ref{m2>0}) are respectively, 
\begin{equation}\label{min00}
\begin{split}
&\left( \lambda_\text{hs}<0 \wedge \mu^2_\text{s} \geq \frac{c_\text{h}\mu^2_\text{h}}{c_\text{h}} \wedge  \lambda_\text{s}> \frac{ \lambda^2_\text{hs}}{ \lambda_\text{s}} \right)\\
&\hspace{1cm}\vee 
 \left(  0<\lambda_\text{hs}<\frac{c_\text{h} \lambda_\text{h}}{ c_\text{h}} \wedge \mu^2_\text{s} \geq \frac{ c_\text{s}\mu^2_\text{h}}{ c_\text{h}} \wedge  \lambda_\text{s}>\frac{ \lambda_\text{hs}\mu^2_\text{s}}{\mu^2_\text{h}}
 \right)\\
 &\hspace{2cm}\vee \left( \lambda_\text{hs} >\frac{c_\text{s} \lambda_\text{h}}{c_\text{h}} \wedge \mu^2_\text{s}>\frac{\lambda_\text{hs}\mu^2_\text{h}}{\lambda_\text{h}} \wedge \lambda_\text{s} > \frac{\lambda_\text{hs} \mu^2_\text{s}}{\mu^2_\text{h}}
 \right)
\end{split}
\end{equation}
and 
\begin{equation}
\begin{split}
 &\left( \lambda_\text{hs}<0 \wedge \mu^2_\text{s} > \frac{\lambda_\text{hs}\mu^2_\text{h}}{\lambda_\text{h}} \wedge  \lambda_\text{s}> \frac{ \lambda^2_\text{hs}}{ \lambda_\text{s}} \right) \\
 &\hspace{1cm}\vee 
 \left(  0<\lambda_\text{hs}<\frac{c_\text{h} \lambda_\text{h}}{ c_\text{h}} \wedge \mu^2_\text{s} > \frac{ \lambda_\text{hs}\mu^2_\text{h}}{ \lambda_\text{h}} \wedge  \lambda_\text{s}>\frac{ \lambda_\text{hs}\mu^2_\text{s}}{\mu^2_\text{h}}
 \right)\,.
 \end{split}
\end{equation}
Eq. (\ref{T2<T1}) is not consistent with Eq. (\ref{min00}), which means the minimum $(0,0)$ cannot coexist again with the minimum $(v,0)$ when we require $T'_c<T_c$. Therefore also this scenario fails to support a first-order phase transition at an intermediate temperature interval. Nevertheless, a second-order phase transition in the first step is possible provided that $\mu^2_\text{h}>0$, $ \lambda_\text{hs}<0$, $ c_\text{s}>0$, $\lambda^2_\text{hs}<\lambda_\text{h}\lambda_\text{s}$ and, 
\begin{equation}
 \left( \mu^2_\text{s}>0 \wedge \lambda_\text{h}>0 \wedge 0<c_\text{h}<\frac{c_\text{s} \mu^2_\text{h}}{\mu^2_\text{s}} \right) ~\vee~  \left( \mu^2_\text{s}<0 \wedge c_\text{h}>0 \wedge 0<\lambda_\text{h}<\frac{\lambda_\text{hs} \mu^2_\text{h}}{\mu^2_\text{s}} \right).
\end{equation}

\begin{table}
 \begin{center}
    \begin{tabular}{ | l | l | l  | p{5cm} |}
    \hline
    Phase transition channel  &  1st-step & 2nd-step   \\ \hline 
    $(0,0)\to(v,0) $  & 2nd-order & \\ \hline  
    $(0,0)\to(v,w)$  &  $\times$ & \\ \hline
    $(0,0)\to(0,w)\to(v,0)$ & 1st- or 2nd-order & 1st-order  \\ \hline
     $(0,0)\to(0,w)\to(v',w')$ & $\times$  & 2nd-order   \\ \hline
      $(0,0)\to(v,w)\to(v',0)$ & $\times$ &  $\times$  \\ \hline
       $(0,0)\to(v,0)\to(v',w')$ & 2nd-order& $\times$  \\ \hline
    \hline
    \end{tabular}
\end{center}
        \caption{All possible electroweak phase transition channels in the $\mathbb{Z}_2$ symmetric singlet scalar extension of the Standard Model.}
        \label{PTtable}
        \end{table}

Note that the vacuum $(v,0)$ is always the global minimum in the temperature interval $0<T'_c<T_c$, because $V(0,0;T)-V(v,0;T)=\mu^4_\text{h}(T)/4\lambda_\text{h}>0$.
In the second step where $0<T<T'_c$ the vacuum solutions $(v,0)$ and $(v',w')$ do not coexist and the first-order phase transition becomes impossible. Also the conditions for the second-order phase transition are in conflict when we require that the phase transition takes place within $0<T<T'_c$. Therefore, in this last scenario only a second-order phase transition in the first step may occur.

\section{Dark Matter and Collider Constraints}\label{secdm}

Because of the $\mathbb{Z}_2$ symmetry in the dark sector, the real singlet scalar can be a Dark Matter (DM) candidate if it takes zero VEV. It is then an interesting question to ask whether or not the electroweak phase transition conditions discussed in the foregoing section are consistent at the same time with the dark matter relic abundance $\Omega\,h^2= 0.120\pm 0.001$ observed by the Planck \cite{Aghanim:2018eyx} and the constraints from DM direct detection experiments such as XENON1T. We answer this question within the framework of the {\it freeze-out} mechanism. 

Among all phase transition channels investigated in the last section, only in scenarios \ref{subsec00v0} and \ref{subsec000wv0}, the singlet scalar has vanishing VEV at lower temperatures possessing the vacuum structure $(v,0)$ in the second step (The scenario \ref{subsec00vwv0} does not lead to a first- or second-order EWPT). 
At $T=0$, the mass eigenvalues for such scenarios are given by, 
\begin{equation}\label{hsmass}
 m^2_h=2\mu^2_\text{h}\hspace{2cm} m^2_s=v^2_\text{h} \lambda_\text{hs} -\mu^2_\text{s}
\end{equation}
where $v_\text{h}=246$ GeV is the VEV of the Higgs. We are interested in combining the DM constraints with the conditions from relevant EWPT channels. 

The thermal evolution of the DM relic abundance is obtained by solving the Boltzman equation,

\begin{equation}
 \frac{dn_s}{dt}=-3Hn_s -\braket{\sigma_\text{ann} v_\text{rel}}\left(n_s^2-(n_s^\text{eq})^2 \right)
\end{equation}
where $\braket{\sigma_\text{ann} v_\text{rel}}$ is the thermal average of the DM annihilation cross section times the relative velocity, and $H$ is the Hubble parameter. 
We take into account also the DM direct detection constraint by the XENON1T/LUX experiments \cite{Aprile:2017iyp,Akerib:2016vxi} which put an upper bound on the DM-nucleon elastic scattering cross section in terms of the DM mass. In the singlet scalar model, the only contribution to the DM-nucleon cross section comes from the $t$-channel Feynman diagram $ss\bar q q$ which in turn is expressed as an effective operator $ss\bar N N$. This interaction gives the DM-nucleon spin-independent elastic scattering, 
\begin{equation}
 \sigma_\text{SI}^\text{N}=\frac{\alpha_N^2\mu_N^2}{\pi m_\text{DM}^2}
\end{equation}
with $\alpha_N$ being the effective coupling and $\mu_N$ being the DM-nucleon reduced mass. 

\begin{figure}
    \includegraphics[width=\linewidth]{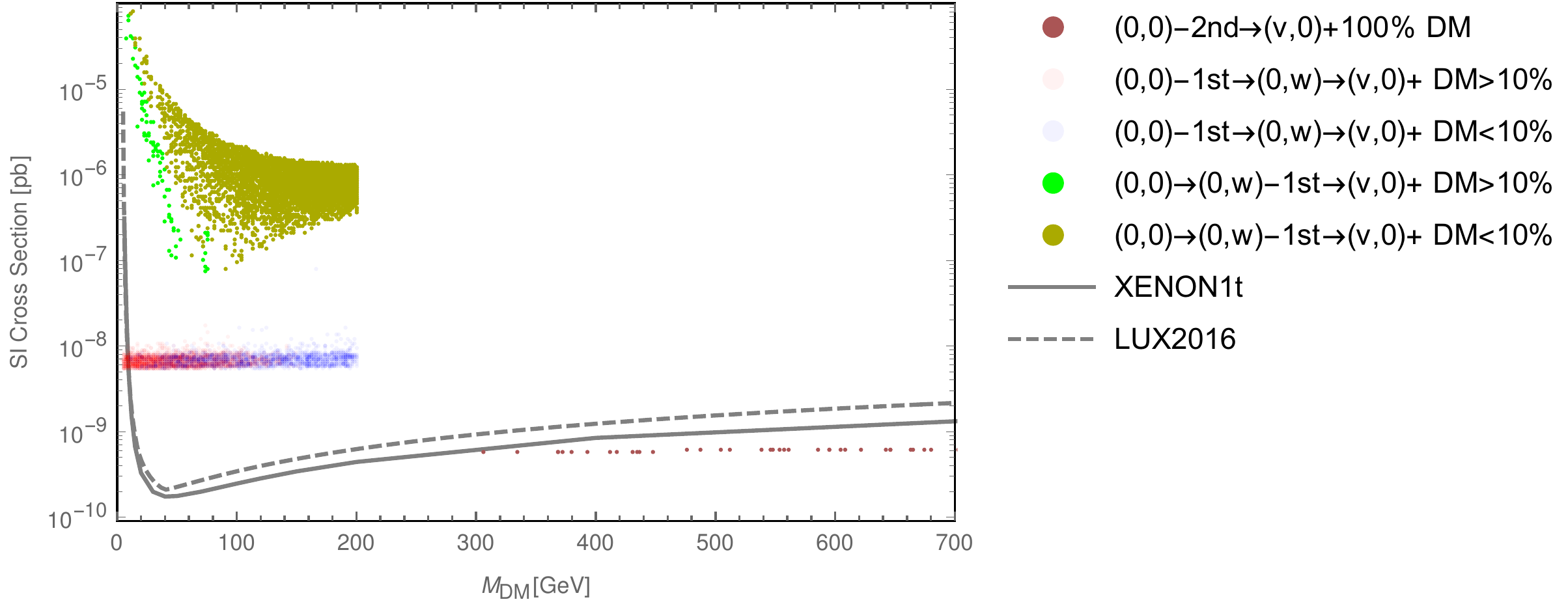}
\caption{The plot shows the DM mass vs. DM-nucleon cross section viable parameter space for one-step and two-step EWPT channels. For two-step EWPT scenarios, the plot is divided into regions where the singlet scalar is responsible for less than or more than $10\%$ of the observed DM relic density. Only the one-step scenarios can evade the XENON1T/LUX direct detection bounds.}\label{dm-xenon1t}
\end{figure}

We calculate both the DM relic density and the DM-nucleon scattering cross section in terms of the DM mass using the {\tt MicroOMEGAs 5.2.7.a} package \cite{Belanger:2010pz}. We take the free parameters to be $m^2_s$, $\lambda_\text{s}$ and $\lambda_\text{hs}$.
In Fig. \ref{dm-xenon1t} we have shown the $(m_\text{DM},\sigma_\text{SI})$ viable parameter space for the EWPT scenarios \ref{subsec00v0} and \ref{subsec000wv0}, divided into regions when the singlet scalar covers less than or more than $10\%$ of the DM relic density.
In the one-step EWPT scenario $(0,0)\to(v,0)$ that can only be second-order, the singlet scalar dark matter is responsible for $100\%$ of the DM relic density for which the DM mass will be in the range $306$ GeV-$5.2$ TeV as seen in Fig. \ref{dm-xenon1t}.  In the two-step EWPT scenario 
$(0,0)\to(0,w)\stackrel{\text{1st}}{\longrightarrow}(v,0)$, the singlet scalar can be responsible for up to about $66\%$ of the DM relic density with the DM mass being in the range $4.5-200$ GeV, while in the EWPT $(0,0)\stackrel{\text{1st}}{\longrightarrow}(0,w)\to(v,0)$ the singlet scalar can explain the whole DM relic density with the DM mass being again $4.5-200$ GeV covering from a fraction up to all DM relic density. The direct detection experiments XENON1T/LUX put strong constraint on the EWPT/DM scenarios; as seen in Fig. \ref{dm-xenon1t}, for both foregoing two-step EWPT, the DM-nucleon scattering cross section is above the cross section bounds found by XENON1T/LUX. However, the one-step EWPT $(0,0)\to(v,0)$ entirely evades the DM-nucleon cross section bounds resulting in a viable DM mass  about $306-5230$ GeV.
It is evident from Fig. \ref{dm-xenon1t} that in two-step EWPT scenarios, the smaller DM mass is, the more percentage of the DM relic density can be explained by the singlet scalar. Also in general, the DM-nucleon cross section in EWPT $(0,0)\stackrel{\text{1st}}{\longrightarrow}(0,w)\to(v,0)$ is at least one order of magnitude smaller than that of $(0,0)\to(0,w)\stackrel{\text{1st}}{\longrightarrow}(v,0)$.

\begin{figure}
    \includegraphics[width=\linewidth]{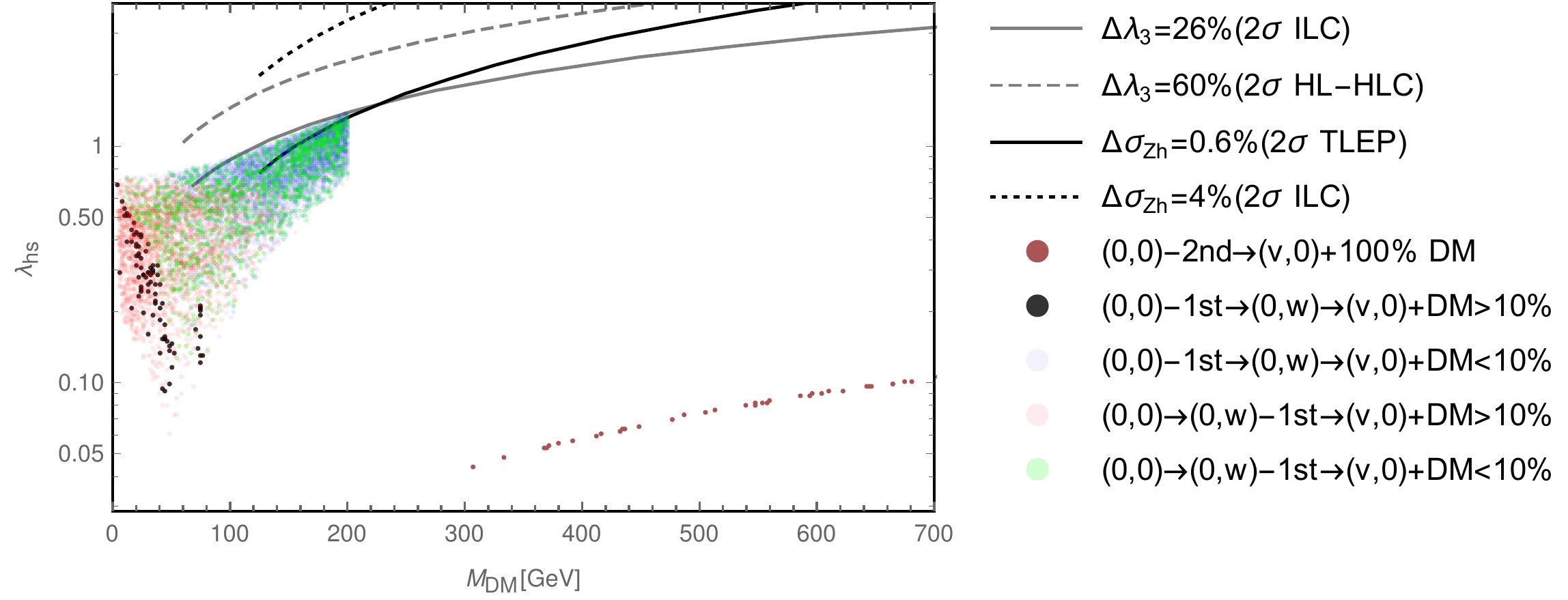}
\caption{The plot shows the viable parameter space in DM mass vs. Higgs portal coupling $\lambda_\text{hs}$ plane, when collider constraints for $\lambda_3$ (triple Higgs coupling) and $\sigma_{Zh}$ ($Zh$ production cross section) are taken into account. The plot is divided into regions where the singlet scalar is responsible for less than or more than $10\%$ of the observed DM relic density. The two-step EWPT scenarios for which singlet scalar covers less than $10\%$ of the DM relic density are accessible to future colliders.}\label{dm-collider}
\end{figure}

It is also interesting to understand if the regions left by imposing the DM constraints in the relevant EWPT scenarios are accessible at colliders. A direct probe of the singlet scalar at the LHC or even a $100$ TeV collider is unlikely \cite{Curtin:2014jma,Beniwal:2017eik}.  However, indirect collider searches are more promising.  
An indirect probe that may hint the presence of the singlet scalar is the triple Higgs coupling. It is obtained from the third derivative of the effective potential at $T=0$ with $\braket{h}=v$ and $\braket{s}=0$ for the EWPT scenarios suitable for the DM candidate, 

\begin{equation}
 \lambda_3=\frac{1}{6}\frac{\partial_3 V_\text{eff}(h,s;T=0)}{\partial h^3}\vert_{h=v_\text{h}}=\frac{m^2_h}{2v_\text{h}}+\frac{\lambda^3_\text{hs} v^3_\text{h}}{24\pi^2 m^2_s}+... \,.
\end{equation}
The best precision in measuring $\lambda_3$ with $2\sigma$ uncertainty is achievable by the $100$ TeV collider to $10\%$ with $30$ ab$^{-1}$ \cite{Barr:2014sga,Contino:2016spe}, at lepton colliders such as ILC $1$ TeV to $13\%$ with $2.5$ ab$^{-1}$ and at the HL-LHC to about $30\%$ \cite{Goertz:2013kp}. Here we make a comparison with the ILC and HL-LHC in Fig. \ref{dm-collider}. It is evident from the Fig. \ref{dm-collider} that the one-step phase transition $(0,0)\stackrel{\text{2nd}}{\longrightarrow} (v,0)$ is not accessible neither at the ILC nor at the HL-LHC due to the smallness of the coupling $\lambda_\text{hs}$. On the other hand, for the two-step phase transitions $(0,0)\stackrel{\text{1st}}{\longrightarrow}(0,w)\to(v,0)$ or $(0,0)\to (0,w)\stackrel{\text{1st}}{\to}(v,0)$ there is a region around $m_s \sim 65$-$100$ GeV which is accessible at the ILC provided that less than $10\%$ of the DM relic density is covered by the singlet scalar.

Another indirect probe at colliders is the contribution of the singlet scalar in $Zh$ production. The correction to $Zh$ production due to the singlet scalar is given by \cite{Englert:2013tya}, 
\begin{equation}
 \Delta \sigma_{Zh}=\frac{\lambda^2_\text{hs} v^2_\text{h}}{8\pi^2 m^2_s}\left( 1+ F(\tau) \right)
\end{equation}
where 
\begin{equation}
 F(\tau)=\frac{1}{4\sqrt{\tau(\tau-1)}}\log \left(\frac{1-2 \tau-2\sqrt{\tau(\tau-1)}}{1-2 \tau+2\sqrt{\tau(\tau-1)}} \right)
\end{equation}
with $\tau=m^2_h/4m^2_s$. The ILC and TLEP programs can probe the $Zh$ cross section to around $2\%$ and $0.3\%$ precision at $1\sigma$ respectively. From Fig. \ref{dm-collider} it is seen that again our one-step phase transition scenario is inaccessible to both colliders. However, both two-step EWPT scenarios $(0,0)\stackrel{\text{1st}}{\longrightarrow}(0,w)\to(v,0)$ and $(0,0)\to (0,w)\stackrel{\text{1st}}{\to}(v,0)$ can be probed by TLEP if $m_\text{DM}\sim 120$-$200$ GeV if the singlet scalar takes only less than $10\%$ of the DM relic density.

\section{Conclusion}
The vacuum structure and phase transition patterns can become very complex with the addition of more scalar degrees of freedom in theories beyond the Standard Model. Among various multi-step phase transition scenarios in a theory with many scalars, one or more first-order phase transitions in an intermediate temperature interval before reaching the zero temperature might be envisioned. In this paper we have studied the vacuum structure and all possible one- and two-step electroweak phase transitions within the high-temperature approximation the most minimal extension to the SM, i.e. the SM with a singlet scalar model with $\mathbb{Z}_2$ discrete symmetry. Although the model is the simplest extension of the SM, there are unexplored phase transition channels that we have studied analytically and in high temperature approximation in this work. In Table \ref{PTtable} a list of all phase transition channels along with the possibility of a first- or second-order phase transition in the first or second step is given.


In this paper we have presented an analytical proof that among all possible phase transition channels in the singlet scalar model, there is only one two-step scenario $(0,0)\to(0,w)\to(w,0)$, that a first-order phase transition can occur,  triggering either from an intermediate critical temperature and finishing at a lower temperature (in the first step) or from a critical temperature down to zero temperature (in the second step). Also two successive first-order phase transition is possible for this channel. 
First-order phase transition in the first step is a new feature for the singlet scalar model. 
For EWPT scenarios discussed in subsections \ref{subsec00v0} and \ref{subsec000wv0}, the singlet scalar takes zero VEV at low temperatures leading to $(v,0)$ vacuum structure at $T=0$. For these scenarios therefore, the singlet scalar is potentially a Dark Matter (DM) candidate. In section \ref{secdm}, we have added the DM and collider constraints on the parameter regions we found for the aforementioned EWPT scenarios. It turns out that for one-step second-order EWPT $(0,0)\to(v,0)$, the singlet scalar can explain the whole observed DM relic density while evading the direct detection bounds from experiments XENON1T/LUX with the DM mass being in the range $300$ GeV-$5.2$ TeV. However, this EWPT scenario is not accessible for future colliders. For the two-step EWPT scenario $(0,0)\to(0,w)\to(v,0)$, if the first-order phase transition occurs in the first step the entire DM relic density if covered by the singlet scalar, but if the first-order takes place in the second step the singlet scalar can explain only a fraction of the DM relic density. Both cases are completely excluded by the DM direct detection experiments XENON1T/LUX, nevertheless at the DM mass around $65-200$ GeV the two-step first-order EWPT scenario can be probed in future colliders through the contribution of the singlet scalar in the triple Higgs coupling and in the $Zh$ production. 
 
 In the EWPT analysis throughout this work we have employed the high-temperature expansion of the thermal effective potential. A complete thermal contribution and the one-loop corrections may change the values of the critical temperature and that in turn may change the nature of the phase transition in different channels.

\section*{Acknowledgement}
I would like to thank José Espinosa, Karim Ghorbani, Alessandro Strumia and Daniele Teresi for discussions. This work was supported by the ERC grant NEO-NAT.


\bibliography{ewpt-singlet-scalar.bib}
\bibliographystyle{unsrt}
\end{document}